\documentclass[graybox]{svmult}

\usepackage{mathptmx}       
\usepackage{helvet}         
\usepackage{courier}        
\usepackage{type1cm}        
\usepackage{makeidx}         
\usepackage{graphicx}        
\usepackage{multicol}        
\usepackage[bottom]{footmisc}

\makeindex             

\begin{document}

\title*{Diffuse light and galaxy interactions in the core of nearby clusters}


\author{Magda Arnaboldi}


\institute{Magda Arnaboldi \at European Southern Observatory, Garching, 
Germany, and Obs. of Turin, INAF, Italy\\
\email{marnabol@eso.org}
}
%
%
\maketitle

\abstract*{The kinematics of the diffuse light in the densest regions of the
nearby clusters can be unmasked using the planetary nebulae (PNs) as
probes of its stellar motions. We describe the results of the most recent 
observations of diffuse light in nearby cluster cores.}

\abstract{The kinematics of the diffuse light in the densest regions
  of the nearby clusters can be unmasked using the planetary nebulae
  (PNs) as probes of the stellar motions. The position-velocity
  diagrams around the brightest cluster galaxies (BCGs) identify the
  relative contributions from the outer halos and the intracluster
  light (ICL), defined as the light radiated by the stars floating in
  the cluster potential.  The kinematics of the ICL can then be used
  to asses the dynamical status of the nearby cluster cores and to
  infer their formation histories. The cores of the Virgo and Coma are
  observed to be far from equilibrium, with mergers currently
  on-going, while the ICL properties in the Fornax and Hydra clusters
  show the presence of sub-components being accreted in their cores,
  but superposed to an otherwise relaxed population of stars. Finally
  the comparison of the observed ICL properties with those predicted
  from $\Lambda$ CDM simulations indicates a qualitative agreement and
  provides insights on the ICL formation. Both observations and
  simulations indicate that BCG halos and ICL are physically distinct
  components, with the ``hotter" ICL dominating at large radial
  distances from the BCGs halos as the latter become progressively
  fainter.}

\section{Diffuse light in clusters}
\label{intro}

\abstract*{A brief overview of the discovery and the follow-up imaging
  studies of diffuse light in clusters.}

Deep imaging of massive clusters of galaxies shows that a population
of star exists that fills the space among galaxies in clusters. While
the indication of its existence came with the early detection by
Zwicky in 1951 for the Coma cluster, it was only with the advent of
the wide-field cameras equipped with mosaic CCDs that the properties
of the clusters' diffuse light, i.e morphology, radial distribution
and colors, were studied in a quantitative way (Uson et al. 1991,
Bernstein et al. 1995, Gregg \& West 1998, Mihos et al. 2005, Rudick
et al. 2010).

Accurate photometric measurements of the diffuse light are difficult
to perform because 1) its features are at extremely faint surface
brightness of $< 1\%$ of the night sky and 2) it is difficult to
disentangle the contribution of the extended outer halos of the
brightest cluster galaxies (BCGs) in a cluster core from that of the
stars free floating in the cluster potential, i.e. the intracluster
light component or ICL. Without the kinematic measurements, the division
between BCG halos and ICL is somewhat arbitrary.

Since the discovery of free-floating intracluster planetary nebulae
(ICPNs) in the Virgo cluster (Arnaboldi et al. 1996), extensive imaging
and spectroscopic observations were carried out to determine their
projected phase space distribution, and from it the amount of ICL.

The measurement of the ICL in clusters is relevant from the
determination of the barionic fraction condensed in stars, the star
formation efficiency and the metal enrichment of the hot intracluster
medium, especially in the cluster cores. Because of the long dynamical
time across the cluster regions, we expect the distribution function
$f({\bar{x},\bar{v}})$ of the ICL stars to be different depending on the
formation mechanism and its assembly history.

\section{PNs as kinematical traces}
\label{PNKT}

\abstract*{This section summarizes the physical properties that make planetary 
nebulae traces of light and kinematics.} 

Planetary Nebulae (PNs) are the late phase of solar-like stars and in
stellar populations older than 2 Gyrs one star every few millions is
expected to undergo such phase (Buzzoni et al. 2006). Stars in the PN
phase can be detected via the relative bright emission in the optical
[OIII] line, at $\lambda$ 5007 \AA, because the nebular shell re-emits
10\% of the UV photons emitted by the white dwarf at its center in the
[OIII] $\lambda$5007 \AA\ line. When such emission line is detected,
the line-of-sight velocity of the PN can be measured via a Gaussian
fit.

The number density of PNs trace the light of the parent stellar
population. According to the single stellar population theory, the
luminosity-specific stellar death rate is independent of the precise
star formation history of the associated stellar population (Renzini
\& Buzzoni 1986). This property is captured in a simple relation such
that
\begin{equation}
N_{PN} = \alpha L_{gal}
\end{equation}
where $N_{PN}$ is the number of all PNs in a stellar population,
$L_{gal}$ is the bolometric luminosity of that parent stellar
population and $\alpha$ is the luminosity-specific PN number.
Predictions from the stellar evolution theory are further supported by
the empirical evidence that the PN number density profiles follow
light in late- and early-type galaxies (Kimberly et al. 2008, Coccato
et al. 2009) and that the luminosity-specific PN number $\alpha$ stays
more or less constant for (B-V) color $< 0.8$, and then decreases by
about a factor 7 for very red (B-V) $> 0.8$ and old stellar population
(Buzzoni et al. 2006).

\section{The Virgo cluster core}\label{Virgo}

\abstract*{This section describes the observations of PNs in the Virgo cluster
core and the implication on its dynamical status.} 

The presence of diffuse light in the Virgo cluster core is clearly
illustrated by the deep image of Mihos et al. (2005), reaching
$\mu_v=28$ mag arcsec$^{-2}$. It shows a variety of features such as
streamers, arcs, plumes and very extended diffuse halos surrounding
the large galaxies in the field at surface brightness level of $\mu_v
\approx 26.5 $ mag arcsec$^{-2}$, and in the case of M87, with
flattened isophotes (c/a=0.5) out to $\sim 37'= 161$ kpc. The
different morphologies suggest that we may be seeing several systems
superposed along the line-of-sight (LOS) to the Virgo core and their
dynamical status may be characterized by a different kinematics.

For the Virgo cluster, there has been considerable success with a
two-step approach of identifying PN candidates with narrow-band
imaging followed by multi-object spectroscopy.  Arnaboldi et
al. (1996) observed the outer regions of the giant elliptical M86,
measuring velocities for 19 objects. While M86's $v_{sys} = - 240$
kms$^{-1}$, three of these PNs have velocities larger than the mean
velocity of the Virgo cluster $\bar{v}_{virgo} = 1100$
kms$^{-1}$, and turned out to be true ICPNs. Subsequently, 23 PNs were
detected in a spectroscopic survey\footnote{These results were all
  based on single line identifications, although the second oxygen
  line was seen with the right ratio in the composite spectrum of 23
  PNs observed by Freeman et al. (2000).} with 2dF on the 4m Anglo-
Australian Telescope (Freeman et al. 2000, Arnaboldi et al. 2002).
The first confirmation based on detecting the [OIII] doublet in a
single PN spectrum was made in Arnaboldi et al. (2003).  Since then,
Arnaboldi et al. (2004, A04) began a campaign to systematically survey
PN candidates in the Virgo cluster using multi-object spectroscopy with
the FLAMES/GIRAFFE spectrograph on the VLT. A04 presented the first
measurements of the velocity distribution of PNs from three survey
fields in the Virgo cluster core and concluded that in two of these
fields the light is dominated by the extended halos of the nearby
giant elliptical galaxies, while the ICL component dominates the
diffuse light in only one field, where a ``broad'' line-of-sight
velocity distribution is measured, and all PNs are true ``ICPNs''.

\begin{figure}[t]
\includegraphics[scale=.5]{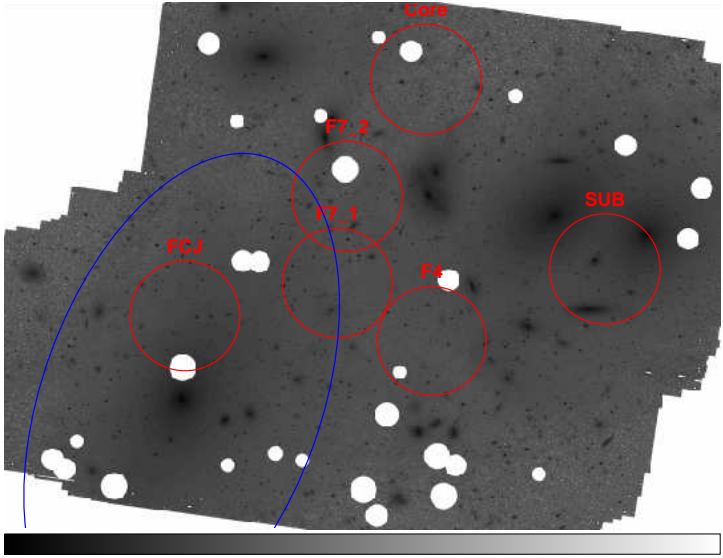}
\caption{Deep image of the Virgo cluster core showing the diffuse
  light distribution (Mihos et al. 2005), with the target fields of
  A04 and D09 indicated by red circles. The blue ellipse shows the
  outer isophotes of the M87 halo according to Kormendy et
  al. (2009).}
\label{fieldsVC}
\end{figure}

The A04 sample was further enlarged with the PN spectra
obtained with FLAMES at the ESO VLT by Doherty et al. (2009, D09), and 
Fig.~\ref{fieldsVC} shows the overview of all the fields in the
Virgo cluster core studied thus far.


\subsection{PNs line-of-sight velocity and projected phase-space distributions
\label{VirgoLOSDV}}

\abstract*{This section describes the properties of the projected
  phase space (PNs $v_{LOS}$ vs. distance from M87) in the Virgo
  cluster core. Three regions are identified: nearby region with (1)
  the densest part associated with the M87 halo, (2) the ICL
  associated with the encroaching stars from the M86/M84 sub-cluster,
  and at larger radii (3) the ICL with a broad velocity distribution
  characteristic of the Virgo cluster.}

On the bases of the PNs' position and $v_{LOS}$ in the Virgo cluster
core, one can computed the projected phase-space diagram by plotting
the PNs $v_{LOS}$ versus radial distance from the center of M87. In
this $v_{LOS}$ vs. $radius$ diagram several regions can be identified
with very different densities: for projected distances $R < 2400''$
most of the PNs are strongly clustered around the systemic velocity of
M87, $v_{sys} = 1307$ kms$^{-1}$, while at $R > 2400''$ the PN
velocities spread widely over a velocity range more typical for the
Virgo cluster. From this intracluster region, we see a string of low
PN velocities at $ \leq 800$ kms$^{-1}$ reaching far into the M87
halo. At projected distance $R < 1300''$ there are two of these
intracluster PNs at $\sim 400$ kms$^{-1}$. The remaining PNs are
distributed symmetrically around M87's $v_{sys}$ and have mean
velocity $1276 \pm 71$ kms$^{-1}$ and velocity dispersion $\sigma =
247$ kms$^{-1}$ (A04).  At $R \sim 2000''$, five PNs are tightly
clustered around $v_{sys} = 1307$ kms$^{-1}$; these have mean velocity
$1297\pm 35$ kms$^{-1}$ and an rms dispersion of $78$ kms$^{-1}$
(D09). At comparable radii there are two additional PNs with
velocities of 753 and 634 kms$^{-1}$; compared to the previous five,
these are 7$\sigma$ and 8 $\sigma$ outliers. It is unlikely that one
or two of these outliers are part of the same (very asymmetric)
distribution as the five PNs clustered around M87's
$v_{sys}$\footnote{The probability of finding 5 PNs with velocity
  dispersion less than 80kms$^{-1}$ from a Gaussian velocity
  distribution with $\sigma = 250$ kms$^{-1}$ around $\bar{v} =
  1300$ kms$^{-1}$ as measured at 60 kpc is less than 1\%.}. By
contrast, they fit naturally into the stream leading from a distance
of $1300''$ from the M87 center all the way into the ICL.


\begin{figure}
\sidecaption
\includegraphics[scale=.35]{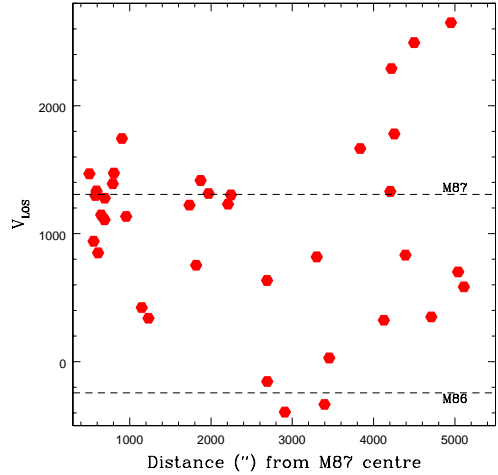}
\caption{Distribution of $v_{LOS}$ versus projected distance from the
  center of M87 for all spectroscopically confirmed PNs in the Virgo
  core. From D09.}
\label{phasesp}
\end{figure}

The PNs bound to the M87 halo are then only those ones which are
clustered around the systemic velocity of M87. These are confined to
radii $R < 2400''$.  Outside $R = 2400''$ in Fig.~\ref{phasesp} the
PNs have larger relative velocities with respect to M87's $v_{sys}$,
with an approximately uniform distribution in the range -300 to -2600
kms$^{-1}$. Those in the radial range $2400''< R < 3600''$ are
confined to negative velocities with respect to M87, indicating that
the ICL component is not phase-mixed yet. These are probably
encroaching stars from M86 and other Virgo components. By contrast,
the PNs further than $3600''$ from M87 show a broad distribution of
velocities, more characteristic of the cluster as a whole.

The ICL PNs show up as approximately flat velocity distribution (VD)
besides the peak of velocities from PNs bound to M87. A flat
distribution of velocities in addition to the peak near M87's systemic
velocity is also seen in the LOSVD of the dwarf spheroidal galaxies in
the same region of the Virgo cluster core (Binggeli et al. 1993).
However, for the dwarf galaxies the broad velocity distribution
extends to significantly more red-shifted velocities, indicating that
the dwarfs and ICL PNs kinematics can only partially be related.


\begin{figure}[t]
\includegraphics[scale=.5]{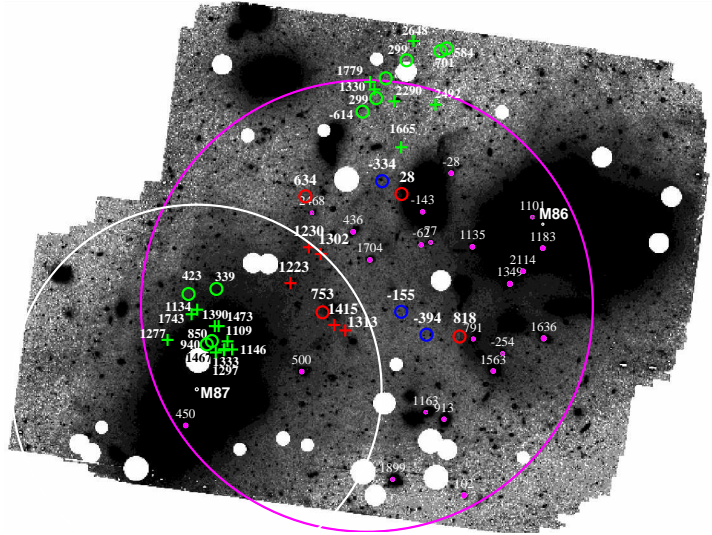}
\caption{Deep image of the Virgo cluster core showing the distribution
  of the intracluster light (from Mihos et al. 2005). The spatial
  distribution of the spectroscopically confirmed PNs are
  overlaid. The A04 targets are shown in green; D09 targets are shown
  in red if red-shifted with respect to Earth and blue if
  blue shifted. Objects with velocities higher than the mean velocity
  of Virgo (1100 km s$^{-1}$) are shown as crosses and those with
  lower velocities shown as circles. Dwarf spheroidals are marked as
  magenta dots. The velocities (in km s$^{-1}$) are labeled for all
  objects shown.  The nominal ``edge'' of the M 87 halo at ${\approx }
  160$ kpc is indicated with a white circle. The pink circle has a 1.5
  degree diameter and is centered on the projected midpoint of M87 and
  M86. North is up and East is to the left. From D09.}
\label{Mihos+PNS}
\end{figure}

In Figure~\ref{Mihos+PNS} we plot the PNs' positions and those of the
dwarfs' in the cluster core. Those PNs in the densest part of the
projected phase space diagram are spatially associated with the halo
of M87, and they are segregated in radii within $R< 161$ kpc, while
the ICPNs are scattered across the whole core region, including the
M87 halo.

\subsection{Dynamical status of the Virgo cluster core}

\abstract*{This section discusses the asymmetry of the LOSVD for ICPNs
  and dwarf galaxies and its implication on the dynamical status of
  the Virgo core. Conclusions from ICPNs and dwarfs LOSVDs are that
  M87 and M86 are falling towards each other nearly along the LOS, and
  in a phase just before their first close pass.}

The LOSVD of dwarf spheroidals (dE+dS0) in a $2^\circ$ radius circular
region centered on M87 is very flat and broad, with the peak of the
distribution at 1300 kms$^{-1}$ and a long tail of negative velocities
(Binggeli et al. 1993). The LOSVD of the ICPNs now confirms that this
asymmetry is also present in the very center of the Virgo core, in a
1$^\circ$ diameter region.  The projected phase-space diagram of
Fig.~\ref{phasesp} shows that velocities near the systemic velocity of
M86 are seen to about half-way from M86 to M87.  The asymmetry and
skewness of the LOSVD may arise from the merging of sub-clusters along
the LOS (Schindler \& Boehringer 1993).  In a merging of two clusters
of unequal mass, the LOSVD is highly asymmetric with a long tail on
one side and a cut-off on the other side, shortly ($\sim10^9$ yr)
before the clusters merge.  The observed LOSVDs of the PNs, GCs (C\^ot\'e
et al. 2001), and (dE+dS0) in the Virgo core can be interpreted as
evidence that the two massive sub-clusters in the Virgo core associated
with the giant ellipticals M87 and M86 are currently falling towards
each other - more or less along the LOS, with M87 falling backwards
from the front and M86 forwards from the back - and will eventually
merge, i.e. the entire core of the Virgo cluster must then be out of
virial equilibrium and dynamically evolving.

What is the relative distance between M87 and M86? Do their halos
already touch each other, or are they just before their close pass?
PNLF distances (Jacoby et al. 1990) and ground-based surface
brightness fluctuation distances indicate that M86 is behind M87 by
just under $\sim 0.15$ mag.  However, the most recent surface
brightness fluctuation measurements find that M87 and M86 are only at
very slightly different distances. Within the errors, the distance
moduli (M87: $31.18\pm0.07$, M86: $31.13\pm0.07$) are consistent with
being either at the same distance or separated by 1-2
Mpc. Unfortunately the evidence from the relative relative distances
of M87/M86 is not conclusive at this stage.

\subsection{ICL Large scale distribution in the Virgo cluster from PNs 
narrow band surveys }

\abstract*{This section presents the results on the ICL distribution
  at large scales in the Virgo cluster. The ICL in Virgo as traced by
  ICPNs is confined to the high density regions: the Virgo core M87
  and M86/M84 over density, plus the M49 and M60/M59 subgroups.}

Several studies investigated the properties of the ICL in the core of
the nearby Virgo cluster by mapping the number density distribution of
ICPNs. Expanding on the earlier narrow band imaging work in the Virgo
cluster core (Arnaboldi et al. 2002, 2003; Feldmeier et al. 1998,
2003, 2004; Okamura et al. 2002; Aguerri et al. 2005), Castro-Rodrigu\'ez et
al. (2009) completed a survey campaign of the ICL distribution at
larger scales, outside the Virgo cluster core. In total, they covered
more than 3 square degrees in Virgo, at eleven different positions in
the cluster and at distances between 80 arcmin and some 100 arcmin
from the Virgo core. In several of these regions, the ICL is at least
two magnitudes fainter than in the core.

The diffuse light observed in the core of a galaxy cluster contains
several luminous stellar components that add up along the LOS to the
cluster center: the extended faint halos of the brightest galaxies and
the ICL contribution. When computing the ICL fraction in the Virgo
core, the surface brightness measurements must then be corrected for
the fraction of stars bound to the extended galaxy halos.  When one
selects only true ICPNs, Castro-Rodrigu\'ez et al. (2009) measure a
surface brightness for the ICL of about $\mu_B = 28.8 - 29.5 $ mag
arcsec$^{-2}$ in the Virgo core and these surface brightness values
are consistent with those inferred from the detection of IC RGB stars
(Williams et al. 2007).

\begin{figure}
\sidecaption
\includegraphics[scale=.5]{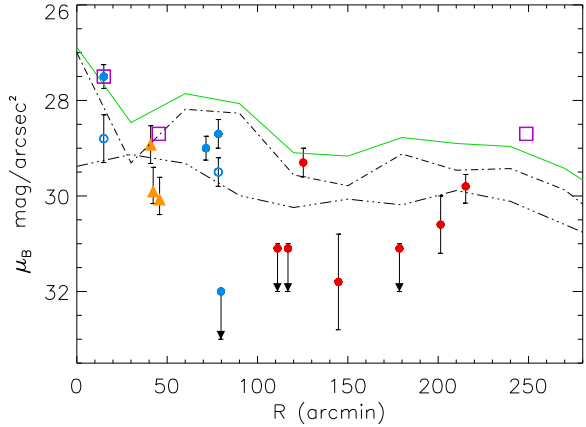}
\caption{Surface brightness measurement of diffuse light in the Virgo
  fields (points) compared with the surface brightness profile of the
  Virgo galaxies averaged in annuli (lines); radial distances are
  computed with respect to M87. The green line represents the radial
  surface brightness profile from light in Virgo galaxies from
  Binggeli et al. (1987).  The dotted-dashed and double dotted-dashed
  lines correspond to the surface brightness profile associated with
  giants and dwarf galaxies, respectively.  The full blue dots show
  the surface brightness measurements in the Virgo core fields using
  ICPNs. The open circles indicate the ICL surface brightness computed
  from true ICPNs, i.e. PNs not bound to galaxy halos. The triangles
  represent the surface brightness of the ICL based on IC RGB star
  counts (see Williams et al. (2007) and the reference therein).  The
  full red dots show the surface brightness measurements outside the
  Virgo core fields using ICPNs. The arrows indicate the upper limits
  for the measurements of PNs at certain field positions. The magenta
  open squares indicate the surface brightness average values $\mu_B$
  at 15, 50 and 240 arcmin based on the measurements from Feldmeier et
  al. (2004) data; the measurements at 240 arcmin (F04-2 and F04-6)
  are close to M49. Radial distances are computed with respect to M87.
  From Castro-Rodrigu\'ez et al. (2009).}
\label{ICLVprof}
\end{figure}

The comprehensive summary of all surface brightness measurements in
Fig.~\ref{ICLVprof} based on ICPN number counts indicates that most of
the diffuse light is detected in fields located in the core, within a
distance of about 80 arcmin, $\approx$ 350 kpc, from M87.  Outside the
core, the mean surface brightness decreases sharply, and the ICL is
confined within isolated pointings.

Diffuse stellar light is also measured in sub-structures, around M49
and in the M60/M59 sub-group. The fields F04-2 and F04-6 from
Feldmeier et al. (2004) are situated in the outer regions of M49, at
about 150 kpc from the galaxy center. These fields may contain PNs
from the halo of M49 and those associated with the ICL component,
which may have formed within sub-clumpB of the Virgo cluster. Because
the spectroscopic follow-up in not yet available for these PN
candidates, we cannot quantify the fraction of light in the M49 halo
and ICL for these fields.

The result for the ICL being more centrally concentrated than the
galaxies in the Virgo cluster is confirmed generally for clusters with
central BCGs. It agrees with statistical observations of intermediate
redshift clusters (Zibetti et al. 2005) and their diffuse light radial
profiles.

\section{Observing techniques for the kinematics of diffuse light in 
clusters \label{obstec}} 

\abstract*{This section contains a short overview of the observing
  techniques that allow studies of the kinematic of the diffuse light
  in clusters far beyond 15 Mpc, and are used to explore distant
  clusters out to Coma, at 100 Mpc.}

In this section we present a short overview of the observing
techniques that allow studies of the kinematic of the diffuse light in
clusters far beyond 15 Mpc, and explore distant clusters out to Coma,
at 100 Mpc. A number of techniques have been developed recently that
allow detection and velocity measurements of PNs in the same observing
run. We refer to the work of M\'endez et al. (2001), which used a
sequence of narrow plus broad band images, followed by the dispersed
image of the same field. This technique has been applied to NGC~4967
in Virgo and to NGC~1344 in Fornax. The PN Spectrograph (PN.S) is an
instrument on the William Herschel Telescope dedicated to measuring PN
velocities in nearby galaxies (Douglas et al. 2002).  The pupil is
split in half before being dispersed in opposite directions in twin
spectrographs. A combination of the two exposures allows the
identification of emission-line objects and their velocity
measurements from the separation between positions in two spectral
images.

\subsection{Counter Dispersed Slitless Imaging technique \label{CDI}}

The counter dispersed slitless technique (CDI) used by McNeil et
al. (2010) uses only two exposures for each field to
obtain positions and velocities of PNs for the cD galaxy NGC~1399, in
the Fornax cluster, at D=17 Mpc distance. The field is observed with a
dispersed image through a narrow band filter centered at the red-shifted
emission of the PN at the systemic velocity of the galaxy being
studied. Next the spectrograph is rotated 180 degrees and the same
field is exposed again, this time with the dispersion in the opposite
direction. As in the PN.S observations, the velocity is a function of
the separation between the position of the PN in the two frames. In
this way, the slitless technique avoids the two stage selection and
measurement process. Because there are no slits or fibers, the light
loss is reduced. The number of measurable PN velocities is not limited
either by the number of available fibers or the restrictive geometry
of the slit. For a comprehensive presentation of the CDI technique and
the calibration procedure to derive relative and absolute velocity,
and sky position of the emission line sources we refer the reader to
McNeil et al. (2010).

\subsection{The Multi-Slit-Imaging-Spectroscopy technique \label{MSIS}}

For the PNs in distant galaxies $D > 20$ Mpc, the flux in the [OIII]
$\lambda$ 5007\AA\ emission becomes of the same order as the sky noise
in a $30-40$ \AA\ wide filter, therefore we need to deploy a technique
which substantially reduced the noise in the background sky to be able
to detect them.  Observing the entire field through slits, a narrow
band filter and a dispersing element significantly reduces the signal
from the sky, because we limit it to few \AA\ only. The Multi slit
Imaging Spectroscopy (MSIS) technique pioneered at the 8.2 m Subaru
telescope and FOCAS spectrograph (Gerhard et al. 2005) involves a grid
of slits that are stepped until the entire field has been spectrally
imaged through a narrow band filter and a grism. It is a blind
technique and spectra are obtained for all the emission line objects
that happen to lie beyond the slits. These type of observations
detected sample of intracluster PNs in the Hydra (Ventimiglia et
al. 2008) and in the Coma cluster (Gerhard et al. 2005, Arnaboldi et
al. 2007).

\section{The un-mixed kinematics of the intracluster stars in the Fornax 
and Hydra cluster cores}

\abstract*{This section provides a concise overview of the most recent
  results on the un-mixed kinematics of intracluster stars in the
  cores of the Fornax and Hydra clusters.}

In this section we provide a concise overview of the most recent
results on the un-mixed kinematics of intracluster stars based on PNs
velocities in the cores of the Fornax and Hydra clusters, from the
works of McNeil et al. (2010) and Ventimiglia et al. (2010, in prep.).

{\it The Fornax cluster and NGC~1399} - Using the CDI technique with
the FORS1 spectrograph at the ESO VLT, Mc. Neil et al. (2010)
discovered 187 PNs around NGC~1399, the cD galaxy in the Fornax
cluster. Data were extracted from a mosaic of 5 fields, which included
also the nearby galaxy NGC~1404. Each PN was further classified on the
basis the light contribution from the two galaxies at its position,
and the difference between the PN velocity and the systemic velocities
of NGC~1399 ($v_{sys} = 1425$ kms$^{-1}$; from NED) and NGC~1404
($v_{sys} = 1917$ kms$^{-1}$; from NED). This procedure identified 146
PN associated with NGC~1399, 23 PNs associated with NGC~1404, while 6
PNs were unassigned. The projected PNs phase space distribution
$v_{LOS}$ vs. radius in the NGC~1399 surveyed regions indicated the
presence of a kinematic component at low velocity $v_{mean} = 800$
kms$^{-1}$ also, in addition to the PN population of NGC~1399 and
NGC~1404. A total of 12 PNs are associated with the low velocity
component, and they are scattered across the NGC~1399 light
distribution, with a concentration in the North-central part. The most
recent globular cluster work also show a sample of low velocity
objects at similar velocity (Schulbert et al. 2010) which
independently support the reality of this substructure.

The presence of a velocity sub-component in the PNs sample superposed
on a PN population bound to the NGC 1399 halo indicate the presence of
a heterogeneous population including stars left over from previous
accretion, merger or tidal stripping events; these stars are still in
the process of phase-mixing.
 
\begin{figure}
\sidecaption
\includegraphics[scale=.4]{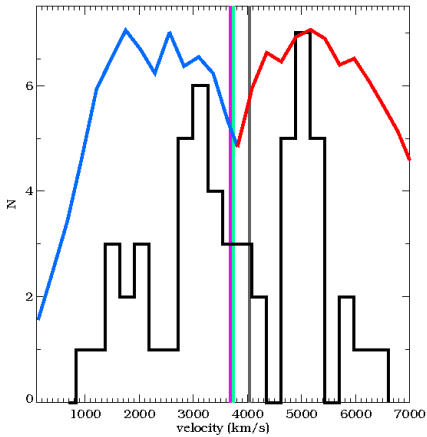}
\caption{The PNs LOSVD in the core of the Hydra cluster obtained from
  the MSIS observations by Ventimiglia et al. (2010). The LOSVD is
  shown by the black line, while the blue and red curves indicate the
  measured transmission curves of the narrow band filters used to
  cover the Hydra velocities; the normalization is arbitrary. The
  magenta, green and gray vertical lines indicate the systemic
  velocity of the Hydra-I cluster, NGC~3311 and NGC~3309,
  respectively.}
\label{NGC3311+PNS}
\end{figure}

{\it The Hydra cluster and NGC~3311} - Ventimiglia et al. (2010, in
prep.) performed MSIS observations with FORS2 at the ESO VLT of the
core region of the Hydra cluster, centered on its cD galaxy NGC~3311.
They detected a total of 56 PNs in a single field of $100 \times 100$
kpc$^2$ , and analyzed their velocity field.  The PNs LOSVD in this
region follows a multi-peaked distribution, see
Figure~\ref{NGC3311+PNS}: in addition to a broad symmetric component
centered at the systemic velocity of the cluster ($v_{hydra} = 3683$
kms$^{-1}$), two narrow peaks are detected at $1800$ kms$^{-1}$ and
$5000$ kms$^{-1}$. These secondary peaks unmask the presence of
un-mixed stellar populations in the Hydra core.

The spatial distribution of the PNs associated with the narrow
velocity sub-component at $5000$ kms$^{-1}$ is superposed and
concentrated on an excess of light in the North-East quadrant of
NGC~3311, as detected from the 2 dimensional light decomposition of
the NGC~3311/NGC~3309 B band photometry. On the same sky position we
detect a group of dwarf galaxies with $v_{LOS} \approx 5000$
kms$^{-1}$.  Deep long-slit spectra obtained at the position of the
dwarf galaxy HCC26 situated at the center of the light excess show
absorption line features from both HCC26 and the light excess which
are consistent with $v_{LOS}$ of about 5000 kms$^{-1}$ (Ventimiglia et
al. 2010). We conclude that the PNs in the $5000$ kms$^{-1}$
sub-component, the dwarfs galaxies and the light excess in the
North-East quadrant of NGC~3311 occupy the same region of the phase
space, and are physically associated.

Also in the case of the relaxed Hydra-I cluster, PN kinematics,
photometry and deep absorption spectra support the evidence for an
accretion event whose stars are being added to both the cluster core
and the halo of NGC~3311.

\section{The on-going sub-cluster merger in the Coma cluster core}

\abstract*{This section provides a short summary of the PNs kinematics
  in the Coma cluster core and the on-going merging of NGC~4889 and
  NGC~4874. }

The Coma cluster is the richest and most compact of the nearby
clusters, yet there is growing evidence that its formation is still
on-going.  A sensitive probe of this evolution is the dynamics of
intracluster stars, which are unbound from galaxies while the cluster
forms, according to cosmological simulations.  With the MSIS technique
Gerhard et al. (2005) detected and measured the $v_{LOS}$ of 37 ICPNs
associated with the diffuse stellar population of stars in the Coma
cluster core, at 100 Mpc distance. These are the most distance singles
stars whose spectra have been acquired in addition to cosmological
supernovae stars.  Gerhard et al.  (2007) detected clear velocity
sub-structures within a 6 arcmin diameter field centered at
$\alpha(J2000) = 12:59:41.8;\, \delta(J2000) 27:53:25.4$, nearly
coincident with the field observed by Bernstein et al. (1995) and
$\sim 5$ arcmin away from the cD galaxy NGC~4874.  A sub-structure is
present at $\sim$ 5000 ${\rm ~km~s^{-1}}$, probably from in-fall of a
galaxy group, while the main intracluster stellar component is
centered around $\sim$6500 ${\rm ~km~s^{-1}}$, $\sim$700 ${\rm
  ~km~s^{-1}}$ offset from the nearby cD galaxy NGC~4874 ($v_{sys} =
7224$ km~s$^{-1}$; from NED) . The kinematics and the elongated
morphology of the intracluster stars (Thuan \& Kormendy 1977) show
that the cluster core is in a highly dynamically evolving state. In
combination with galaxy redshift and X-ray data, this argues strongly
that the cluster is currently in the midst of a sub-cluster merger.
The NGC~4889 sub-cluster is likely to have fallen into Coma from the
eastern A2199 filament, in a direction nearly in the plane of the sky,
meeting the NGC~4874 sub-cluster arriving from the west. The two inner
sub-cluster cores are presently beyond their first and second close
passage, during which the elongated distribution of diffuse light has
been created.  Gerhard et al.  (2007) also predict the kinematic
signature expected in this scenario, and argue that the extended
western X-ray arc recently discovered traces the arc shock generated
by the collision between the two sub-cluster gas halos.

\section{Cosmological simulations and ICL \label{CSICL}}

\abstract*{This section provide a summary of the ICL properties in
  $\Lambda$ CDM cosmological simulations.}

The predicted spatially averaged radial distribution of ICL from
recent high resolution hydrodynamical simulations of cluster formation
in $\Lambda$ CDM universe (Murante et al. 2004) is in broad agreement
with the observed radial profiles for the ICL in clusters.
Furthermore predictions from these simulations that the largest
portion of the ICL is formed during the assembly of the most luminous
cluster galaxies (Rudick et al. 2006, Murante et al. 2007) is
supported by the observed ongoing mergers in Coma and Virgo cores, and
by the presence of ICL around sub-clusters/groups.

Quantitative analysis of the ICL kinematic from cosmological
simulations is on-going (Coccato et al. 2010, in prep). Studies of the
galaxy halo and ICL particles in cosmological simulations (Dolag et
al. 2010) further support the physical distinction between the central
BCG and the ICL component in clusters.

\section{Summary and Conclusions}

\abstract*{The superposition of the various kinematic components
  underscores the complexity of velocity measurements in cluster cores
  and the necessity of using discrete tracers to detect these
  components. The evidence for merging and accretion indicates that the
  build up of the ICL is an on-going process.}

The kinematics of the ICL provide unique information to asses the
dynamical status of the nearby cluster cores:
\begin{itemize} 
\item In the Virgo cluster, M87 and the M86/M84 sub-clusters are
  approaching along the LOS and they are currently before their first
  close passage. 
\item In the Coma cluster, the elongated morphology and kinematics of
  the ICL, the galaxy morphology and X-ray data are consistent with
  the merging of NGC~4889 and NGC~4874 along a binary orbit, the two
  galaxies being currently observed beyond their first and second
  close passage.
\item the Hydra and Fornax clusters show evidence for un-mixed stellar
  components coming from accretion/tidal stripping events in their
  cores.
\end{itemize}

The superposition of the various kinematic components underscores the
complexity of velocity measurements in cluster cores and the necessity
of using discrete tracers to detect these components.  In all cases
the kinematical data indicate that the galaxy halos and ICL are
discrete components; and the former do not blend continuously in the
latter. The evidence for merging (in Coma) and accretion indicates
that the build up of the ICL is an on-going process.

\begin{acknowledgement}
I wish to acknowledge my long standing and productive collaboration
with Ken Freeman, which lead to some forty two published refereed
papers and tens of observing nights carried out together at several
telescopes around the world (2.3 m SSO, AAT, CFHT, VLT, and Subaru to
name a few). Dear Ken, I look forward to our future collaborations!

I also thank David Block for the invitation to participate at the
conference to honor Ken Freeman and all his hard work, dedication
and endurance which made the Ken Freeman Conference in the Namibia
desert and its proceedings come true.

Finally I wish to thank my collaborators Ortwin Gerhard, Lodovico
Coccato, Payel Das, Michelle Doherty, Emily Mc Neil, Nieves
Castro-Rodrigu\'ez, and Giulia Ventimiglia.
\end{acknowledgement}
%

%
%

\begin{thebibliography}{99.}%
\bibitem[]{aguerri+05} Aguerri, M., et al. 2005, 
AJ, 457, 771
\bibitem[]{Arnaboldi+96} Arnaboldi, M., et al. 1996, 
ApJ, 472, 145
\bibitem[]{Arnaboldi+02} Arnaboldi, M., et al. 2002,
AJ, 123, 760
\bibitem[]{Arnaboldi+03} Arnaboldi, M., et al. 2003, 
AJ, 125, 514
\bibitem[]{Arnaboldi+04} Arnaboldi, M., et al. 2004, 
ApJ, 614, L33 (A04)
\bibitem[]{Arnaboldi+07} Arnaboldi, M., et al. 2007, 
PASJ, 59, 419
\bibitem[]{Berstein+95} Bernstein, G.M., et al. 1995, AJ, 110, 1507
\bibitem[]{Binggeli+87} Binggeli, B., Tammann
  G.A., Sandage, A. 1987, AJ, 94, 251
\bibitem[]{Binggeli+93} Binggeli, B., Popescu,
  C.C., Tammann, G.A. 1993, A\&AS, 98, 275
\bibitem[]{BAC06} Buzzoni, A., Arnaboldi, M.,
    \& Corradi, R.L.M. 2006, MNRAS, 368, 877
\bibitem[]{castro+09} Castro-Rodrigu\'ez, N., et al. 2009, A\&A, 507, 621
\bibitem[]{coccato+09} Coccato, L. et al. 2009, MNRAS, 394, 1249
\bibitem[]{Cote+01} C\^ot\'e, P., et al. 2001, ApJ, 559, 828
\bibitem[]{Doherty+09} Doherty, M., et al. 2009, A\&A, 502, 771 (D09)
\bibitem[]{Dolag+10} Dolag, K. et al. 2010, MNRAS, 405, 1544
\bibitem[]{Douglas+02} Douglas, N., et al. 2002, PASP, 114, 1234
\bibitem[]{feldm+98} Feldmeier, J.J., et al. 1998, ApJ, 503, 109
\bibitem[]{feldm+03} Feldmeier, J.J., et al. 2003, ApJS, 145, 65
\bibitem[]{feldm+04} Feldmeier, J.J., et al. 2004, ApJ, 615, 196
\bibitem[]{Freeman+00} Freeman, K.C., et al. 2000,  ASP Conf. Ser., 197, 389
\bibitem[]{OG+05 } Gerhard, O., et al. 2005, ApJL, 621, 93
\bibitem[]{OG+07 } Gerhard, O., et al. 2007, A\&A, 468, 815
\bibitem[]{GWest+98 } Gregg, M.D., \& West, M., 1998, Nature, 396, 549
\bibitem[]{Jacoby90} Jacoby, G., et al. 1990, ApJ, 356, 332
\bibitem[]{Kormendy+09} Kormendy, J. et al. 2009, ApJS, 182, 216
\bibitem[]{kimciar08} Kimberly, H.A, et al. 2008, ApJ, 683, 630
\bibitem[]{Mcneil+10} McNeil, E.K., et al. 2010, A\&A, 518, 44
\bibitem[]{Mendez+01} M\'endez, R.H., et al. 2001, ApJ, 563, 135
\bibitem[]{Mihos+05} Mihos, J.C., et al. 2005, ApJL, 631, 41
\bibitem[]{Murante+04} Murante, G., et al. 2004, ApJL, 607, 83
\bibitem[]{Murante+07} Murante, G., et al. 2007, MNRAS, 377, 2
\bibitem[]{okamura+02} Okamura, S., et al. 2002, PASJ, 54, 883
\bibitem[]{RB86} Renzini, A. , Buzzoni, A., 1986, in Chiosi C.,
  Renzini A., eds., Spectral Evolution of Galaxies. Reidel, Dordrecht,
  p. 195
\bibitem[]{Rudick+06} Rudick, C.S., et al. 2006, ApJ, 648, 936
\bibitem[]{Rudick+10} Rudick, C.S., et al. 2010, ApJ, 720, 569
\bibitem[]{schiboh93} Schindler, S., \&  Boehringer, H. 1993, A\&A, 269, 83
\bibitem[]{schubert+10} Schuberth, Y., et al. 2010, A\&A, 513, 52
\bibitem[]{tk1977} Thuan, T.X., \& Kormendy, J. 1977, PASP, 89, 466
\bibitem[]{uson+91} Uson, J.M., et al. 1991, ApJ, 369, 46
\bibitem[]{vent+08} Ventimiglia, G., Arnaboldi, M., \& Gerhard,
  O. 2008, Astronomische Nachrichten, MNRAS, 329, 1057
\bibitem[]{Zibetti+05} Zibetti, S., et al. 2005, MNRAS, 358, 949
\bibitem[]{Williams+07} Williams, B.F., et al. 2007, ApJ, 654, 835

\end{thebibliography}
%

\end{document}